\documentclass[10pt]{amsart}

\newtheorem{theorem}{Theorem}[section]
\newtheorem{lemma}[theorem]{Lemma}  
\newtheorem{proposition}[theorem]{Proposition}
\newtheorem{corollary}[theorem]{Corollary}
\theoremstyle{definition}
\newtheorem{definition}[theorem]{Definition}
\newtheorem{example}[theorem]{Example}

\theoremstyle{remark}
\newtheorem{remark}[theorem]{Remark}
\newtheorem{question}[theorem]{Question} 

\numberwithin{equation}{section}

\newcommand{\cal}{\mathcal}

\newcommand{\Id}{\text{Id}}
\newcommand{\C}{{\bf C}}

\renewcommand{\H}{{\bf H}}
\newcommand{\M}{{\bf M}}
\newcommand{\1}{{\bf 1}}

\newcommand{\ran}{\text{ran}}
\newcommand{\Comp}{{\text{Comp}}}

\begin{document}

\title{Compact Orthoalgebras}

\author{Alexander Wilce}
\address{Department of Mathematics, Susquehanna University, Selinsgrove, PA 17870}
\email{wilce@susqu.edu}

\subjclass{Primary 06F15, 06F30; Secondary 03G12, 81P10}

\keywords{orthoalgebra, effect algebra, orthomodular lattice, topological lattice, quantum logic }

\begin{abstract} We initiate a study of topological orthoalgebras (TOAs), concentrating on the compact case. Examples of TOAs include topological orthomodular lattices, and also the projection lattice of a Hilbert space. As the latter example illustrates, a lattice-ordered TOA need not be a topological lattice. However, we show that a compact Boolean TOA is a topological Boolean algebra. Using this, we prove that any compact regular TOA is atomic, and has a compact center. We prove also that any compact TOA with isolated $0$ is of finite height. We then focus on {\em stably ordered} TOAs: those those in which the upper-set generated by an open set is open. These include both topological orthomodular lattices and interval orthoalgebras, in particular projection lattices. We show that the topology of a compact stably-ordered TOA with isolated $0$ is determined by that of of its space of atoms. \end{abstract}

\maketitle 
  
\section{Introduction}\label{intro}

Broadly speaking, a {\em quantum logic} is any of a range of order-theoretic 
and partial-algebraic structures -- orthomodular lattices and posets, orthoalgebras, and effect algebras -- abstracted from the projection lattice $L(\H)$ of a Hilbert space $\H$.
Since the primordial example is very much a topological object, it would seem natural 
to undertake a study of ``topological quantum logics" more generally. There does exist a literature devoted to topological orthomodular lattices (e.g., \cite{ChoGre,ChoGreCha,PulRie}); however $L(\H)$, in its norm or strong operator topology, is not a topological lattice, the meet and join in $L(\H)$ being not continuous. On the other hand, $L(\H)$ {\em is} a topological orthoalgebra in a natural sense -- as, indeed, are many other orthoalgebras one meets in practice, including all topological orthomodular lattices. 

The purpose of this paper is to begin a systematic study of 
topological orthoalgebras (TOAs) {\em in abstracto}. In the interest of 
making what follows self-contained, section 2 collects some general 
background material on orthoalgebras. Section 3 develops some of the general 
theory of TOAs, with a focus on the compact case. Among other things, it is 
shown that a compact Boolean TOA is a topological Boolean algebra. This is a 
non-trivial fact, since, as the example of $L(\H)$ shows, a lattice-ordered 
TOA need not be a topological lattice. We also show that any (algebraically) 
regular compact topological orthomodular poset is atomic, and that a compact 
TOA with $0$ isolated is atomic and of finite height.   In section 4, we 
consider a class of TOAs we call {\em stably ordered}: those in which the 
upper-set generated by an open set is again open. This includes all 
topological orthomodular lattices and also projection lattices. We show that the topology of a stably-ordered TOA with $0$ isolated is entirely determined by that on its space of atoms. 

\section{Background}\label{an apprpriate label} 
  
If $(L,\leq,',0,1)$ is any orthocomplemented poset, we call elements $a$ and $b$ of $L$
{\em orthogonal}, writing $a \perp b$, iff $a \leq b'$. Suppose that 
any two orthogonal elements of $L$ have a join. Then for $a \leq b$ in $L$, we can define a relative complement 
$b \wedge a' \perp a$. $L$ is an {\em orthomodular poset} (hereafter: OMP) iff, in addition, 
\begin{equation} 
a \leq b \ \Rightarrow \ (b \wedge a') \vee a = b.\end{equation}
An {\em orthomodular lattice} (OML) is a lattice-ordered OMP. Evidently, the ``orthomodular identity" (1) is a weak form of distributivity, and thus every Boolean algebra is an OML. The primordial (non-Boolean) example is the lattice $L(\H)$ of projections -- equivalently, closed subspaces -- of a Hilbert space $\H$. Orthomodular lattices and posets have been studied extensively. The standard reference is \cite{Kalmbach}; for a more recent survey, see \cite{BruHar}

Let us agree to write $a \oplus b$ for the join of orthogonal elements $a$ and $b$ of an orthocomplemented poset, whenever this join exists. It is not difficult to check that $L$ is an OMP iff the resulting structure $(L,\oplus)$ satisfies the conditions that (i) $a \oplus b$ exists whenever $a \perp b$, and (ii) if $a \oplus b = 1$, then $b = a'$. This 
suggests the following.

\begin{definition}  An {\em orthoalgebra} is a structure 
$(L,\oplus)$ consisting of a set $L$, an associative, commutative\footnote{The  associativity and commutativity of $\oplus$ 
are here to be understood in
the strong sense, i.e., if $a \oplus b$ is defined, then so is $b \oplus a$,
and the two are equal, and if $a \oplus (b \oplus c)$ is defined, so is 
$(a \oplus b) \oplus c$, and the two are equal.} partial binary operation $\oplus$ on $L$, such that for all $a \in L$
\begin{enumerate}
\item[(a)] there exists a unique element $a' \in L$ with $a \oplus a' = 1$;
\item[(b)] $a \oplus a$ exists only if $a = 1'$.\end{enumerate} 
We write $a \perp b$ to indicate that $a \oplus b$ exists. Also, we write $0$ for $1'$, noticing that $0 \oplus a = a$ for every $a \in L$.\end{definition}

Orthoalgebras were introduced in the early 1980s by D. J. Foulis and C. H. Randall [7] in connection with the problem of defining tensor products of quantum logics. Further information can be found in [5] and [16]. For later reference, we mention that an {\em effect algebra} (see, e.g., [1]) is a structure $(L,\oplus)$ satisfying condition (a) and, in place of (b), the weaker condition that for all $a \in L$, if $a \perp 1$, then 
$a = 0$. 

\subsection{Orthoalgebras as orthoposets} From the remarks preceding Definition 1.1, it is clear that any OMP gives rise to an orthoalgebra in which $a \oplus b = a \vee b$. Any orthoalgebra $(L,\oplus)$ can be partially ordered by setting $a \leq b$ iff there exists $c \in L$ with $b = a \oplus c$. The
operation $a \mapsto a'$ is an orthocomplementation with respect to this ordering. Thus, any orthoalgebra gives rise to an orthoposet. Moreover, for
any $a \leq b$ in $L$, there is a unique element $c \in L$ -- namely,
$(b \oplus a')'$ -- such that $b = a \oplus c$. It is usual to call this
element $b \ominus a$. If $L$ is an OMP, this is exactly $b \wedge a'$. In this language, 
the orthomodular law (1) becomes 
\begin{equation}
a \leq b \ \Rightarrow \ b = (b \ominus a) \oplus a, \end{equation}
which holds in any orthoalgebra. In general, however, $a \oplus b$ is not the join, but only a {\em minimal} upper bound, for orthogonal elements $a$ and $b$ of an orthoalgebra $L$. Indeed, one can show that the orthoposet $(L,\leq,',0,1)$ obtained from $(L,\oplus)$ is an OMP if and only if $a \oplus b = a \vee b$ for all $a, b \in L$; this in turn is equivalent to the condition, called {\em orthocoherence} in the literature, that if $a,b,c \in L$ are pairwise orthogonal, then $a \perp (b \oplus c)$, so that $a \oplus (b \oplus c)$ exists. Thus, orthomodular posets are effectively the same thing as orthocoherent orthoalgebras, and orthomodular lattices are effectively the same things as lattice-ordered orthoalgebras. 

\subsection{Boolean orthoalgebras and Compatibility} 
An orthoalgebra $(L,\oplus)$ is said to be {\em Boolean} iff the corresponding orthoposet $(L,\leq,',0,1)$ is a Boolean lattice. 
A subset of $L$ is said to be {\em compatible} iff it is contained in a Boolean sub-orthoalgebra of $L$. Two elements $a, b \in L$ are compatible iff there exist elements $a_1, b_1$ and $c$ with $a = a_1 \oplus c$, $b = c \oplus b_1$, and $a \perp b_1$, so that $a_1 \oplus c \oplus b_1$ exists [5]. Equivalently, $a$ and $b$ are compatible iff there exists an element $c \leq a, b$ with $a \perp (b \ominus c)$. The triple $(a_1, c, b_1) = (a \ominus c, c, b \ominus c)$ is then called a {\em Mackey decomposition} for $a$ and $b$. If $L$ is Boolean, then every pair of elements $a, b \in L$ has a {\em unique} Mackey decomposition, namely, $(a \ominus b, a \wedge b, b \ominus a)$.  It is possible, even in an OMP, for  a pairwise compatible set of elements not to be compatible. An orthoalgebra in which pairwise compatible sets {\em are} compatible is said to be {\em regular}. 

\subsection{The center of an orthoalgebra} 
For any $a \in L$ there is a natural mapping $[0,a] \times [0,a'] \rightarrow L$ given by $(x,y) \mapsto x \oplus y$. If this mapping is in fact an isomorphism, $a$ is said to be {\em central}. The {\em center} of $L$ is the set $\C(L)$ of all central elements of $L$.  It can be shown [8] that $\C(L)$ is a Boolean sub-orthoalgebra of $L$. In particular, $L$ is Boolean iff $L = \C(L)$. We shall call $L$ {\em simple} iff $\C(L) = \{0,1\}$. 

\subsection{Joint orthogonality} A compatible pairwise-orthogonal set is said to be {\em jointly orthogonal}. Equivalently, $A \subseteq L$ is jointly orthogonal iff, for every finite subset $F = \{a_1,...,a_n\} \subseteq A$, the ``partial sum" $\bigoplus F = a_1 \oplus \cdots \oplus a_n$ exists. If the join of all partial sums of $A$ exists, we denote it by $\bigoplus A$, and speak of this as the sum of $A$. We shall say that $L$ is 
{\em orthocomplete} if every jointly orthogonal subset of $L$ has a sum in this sense. An orthoalgebra is {\em atomic} iff every element of $L$ can be expressed as the sum of a jointly orthogonal set of atoms.

\section{Topological Orthoalgebras}
 
\begin{definition} A {\sl topological orthoalgebra} (hereafter: TOA) is an
orthoalgebra $(L,\oplus)$ equipped with a topology making the relation
$\perp \subseteq L \times L$ closed, and the mappings $\oplus : \perp
\rightarrow L$ and $' : L \rightarrow L$, continuous.\end{definition}

One could define a topological effect algebra in just the same way. We shall not pursue this further, except to note that the following would carry over verbatim to that context: 

\begin{lemma}{ Let $(L,\oplus)$ be a topological
orthoalgebra. Then
\begin{enumerate}
\item[(a)] The order relation $\leq$ is closed in $L \times L$
\item[(b)] $L$ is a Hausdorff space.
\item[(c)] The mapping $\ominus : \leq \rightarrow L$ is
continuous.                                           \end{enumerate}}\end{lemma}

\begin{proof} For (a), notice that $a \leq b$ iff $a \perp b'$. Thus, $\leq = f^{-
1}(\perp)$ where $f : L \times L \rightarrow L \times L$ is the continuous
mapping $f(a,b) = (a,b')$. Since $\perp$ is closed, so is $\leq$. That $L$ is
Hausdorff now follows by standard arguments (cf. [9, Ch. VII] or [12]). Finally, since
$b \ominus a = (b \oplus a')'$, and $\oplus$ and $'$ are both continuous,
$\ominus$ is also continuous. \end{proof}

\subsection{Examples} 
Any product of discrete orthoalgebras, with the product topology, is a TOA. Another source of examples is are topological orthomodular lattices (TOMLs) \cite{ChoGre,ChoGreCha}. A TOML is an orthomodular lattice equiped with a Hausdorff topology making the lattice operations, and also the orthocomplementation, continuous. If $L$ is a TOML and $a,b \in L$, then $a \perp b$ iff  $a \leq b'$ iff $a = a\wedge b'$. This is obviously a
closed relation, since $L$ is Hausdorff and $\wedge, '$ are continuous. Thus, every 
TOML may be regarded as a TOA. However, there are simple and important examples of lattice-ordered TOAs that are not TOMLs:

\begin{example} $L$ be the horizontal sum of four-element Boolean algebas
$L_{x} = \{0,x,x',1\}$ with $x$ (and hence, $x'$) parametrized by a non-degenerate real
interval $[a,b]$. Topologize this as two disjoint copies, $I$ and $I'$, of $[a,b]$ plus
two isolated points $0$ and $1$: Then the orthogonality relation is obviously
closed, and $\oplus$ is obviously continuous; however, if we let $x
\rightarrow x_o$ (with $x \not = x_{o})$ in $I$, then we have $x \wedge x_{o}
= 0$ yet $x_o \wedge x_{o} = x_{o}$; hence, $\wedge$ is not continuous. \end{example}

\begin{example} Let $\H$ be a Hilbert space, and let $L = L(\H)$ be the
space of projection operators on $\H$, with the operator-norm topology. The relation
Since multiplication is continuous, the relation $P \perp Q$ iff $PQ = QP = 0$ is closed;
since addition and subtraction are continuous, the partial operation $P, Q \mapsto P \oplus Q := P+Q$ is
continuous on $\perp$, as is the operation $P \mapsto P' := \1 - P$. So $L(\H)$ is a lattice-ordered
topological orthoalgebra. It is not, however, a topological lattice. Indeed, if $Q$ is a non-trivial
 projection, choose unit vectors $x_n$ not
lying in $\ran(Q)$ that converge to a unit vector in $x \in \ran(Q)$. If $P_n$ is the projection generated by $x_n$ and $P$, that generated by with $x$, then $P_n \rightarrow P$. But
$P_n \wedge Q = 0$, while $P \wedge Q = P$.
\end{example}

\begin{remark} Topologically, projection lattices and TOMLs are strikingly different. Any compact TOML is totally disconnected [4, Lemma 3]. In strong contrast to this, of $\H$ is finite dimensional, then $L(\H)$ is compact, but the set of projections of a given dimension in $L(\H)$ is a manifold.  As this illustrates, TOAs are much freer objects topologically than TOMLs. Indeed, by an easy generalization of Example 3.3, any Hausdorff space can be embedded in a TOA. \end{remark}



\subsection{Compact Orthoalgebras} For the balance of this paper, we concentrate on compact TOAs. It is a standard fact [9, Corollary VII.1.3] that any ordered topological space with a closed order is isomorphic to a closed subspace of a cartesian power of $[0,1]$ in its product order and topology. It follows that such a space $L$ is {\em topologically order-complete}, meaning that any upwardly-directed net in $L$ has a supremum, to which it converges.  Applied to a compact TOA, this yields the following useful completeness result:

\begin{lemma} Any compact TOA $L$ is orthocomplete. Moreover, if $A \subseteq L$ is jointly orthogonal, the net of finite partial sums of $A$ converges topologically to $\bigoplus A$. \end{lemma} 

We are going to show that any compact regular TOA is atomic. In aid of this, the following technical definition proves most useful: 

\begin{definition} If $L$ is any orthoalgebra, let 
\[\M(L) := \{ (a,c,b) \in L \times L \times L | c \leq a, \ c \leq b, \ \text{and} \ a \perp (b \ominus c)\}.\] 
In other words, $(a,c,b) \in \M(L)$ iff $(a \ominus c, c, b \ominus c)$ is a Mackey decomposition for $a$ and $b$. \end{definition}

\begin{lemma} For any TOA $L$, the relation $\M(L)$ is closed in $L\times L \times L$.\end{lemma}

\begin{proof} Just note that
$\M(L) = (\geq \times L) \cap (L \times \leq) \cap (\Id \times \ominus)^{-1}(\perp)$.
Since the relations $\leq$ and $\perp$ are closed and $\ominus : \leq \rightarrow L$ is continuous, this also is closed. \end{proof}

Since lattice-ordered TOAs need not be topological lattices, the following is noteworthy:    

\begin{proposition} A compact Boolean topological orthoalgebra is a
topological lattice, and hence, a compact topological Boolean algebra.
\end{proposition} 

\begin{proof} If $L$ is Boolean, then $\M(L)$ is, up to a permutation, the
graph of the mapping $a,b \mapsto a \wedge b$. Thus, by Lemma 3.8,
$\wedge$ has a closed graph. Since $L$ is
compact, this suffices to show that $\wedge$ is continuous.\footnote{Recall
here that if $X$ and $Y$ are compact and the graph $G_f$ of $f : X \rightarrow Y$ is
closed, then $f$ is continuous. Indeed, let $F \subseteq Y$ be closed. Then
$f^{-1}(F) = \pi_{1}((X \times F) \cap G_{f})$, where $\pi_{1}$ is projection
on the first factor. Since $X$ and $Y$ are compact, $\pi_{1}$ sends closed
sets to closed sets.} It now follows from the continuity of $'$ that $\vee$
is also continuous. \end{proof}

Note that every compact topological Boolean algebra has the form $2^{E}$, where $E$ is a set and $2^{E}$ has the product topology \cite{Johnstone}. In particular, every compact Boolean algebra is atomic. This will be useful below.

\begin{question} Is every Boolean TOA a topological Boolean algebra? \end{question}

For any orthoalgebra $L$, let $\Comp(L)$ be the set of all
compatible pairs in $L$, and for any fixed $a \in L$, let $\Comp(a)$ be the set
of elements compatible with $a$. 

\begin{proposition} Let $L$ be a compact TOA. Then
\begin{enumerate}
\item[(a)] $\Comp(L)$ is closed in $L \times L$;
\item[(b)] For every $b \in L$, $\Comp(b)$ is closed in $L$;
\item[(c)] The closure of a pairwise compatible set in $L$ is pairwise compatible;
\item[(d)] A maximal pairwise compatible set in $L$ is closed.
\end{enumerate}\end{proposition}

\begin{proof} (a) $\Comp(L) = (\pi_{1} \times \pi_{3})(\M(L))$. Since $\M(L)$ is closed,
and hence compact, and $\pi_{1} \times \pi_{3}$ is continuous, $\Comp(L)$ is
also compact, hence closed. For (b), note that $\Comp(b) = \pi_{1}(\Comp(L) \cap
(L \times \{b\}))$. Since $\Comp(L)$ is closed, so is $\Comp(L) \cap (L \times
\{b\})$; hence, its image under $\pi_{1}$ is also closed (remembering
here that $L$ is compact). For (c), suppose $M \subseteq L$ is pairwise
compatible. Then $M \times M \subset \Comp(L)$. By part (a), $\Comp(L)$ is closed,
so we have
\[\overline{M} \times \overline{M} \subseteq \overline{M \times M} \subseteq
\Comp(L),\]
whence, $\overline{M}$ is again pairwise compatible. Finally, for (d), if $M$ is
a maximal pairwise compatible set, then the fact that $M \subseteq
\overline{M}$ and $\overline{M}$ is also pairwise compatible entails that
$M = \overline{M}$. \end{proof}

There exist (non-orthocoherent) orthoalgebras in which $\Comp(L) = L \times L$ (\cite{FouGreRut}, Example 3.5). However, in an OML, $\Comp(L) = \C(L)$, the center of $L$. Thus we recover from part (a) of Proposition 3.11 the fact (not hard to prove directly; see \cite{ChoGre}) that the center of a compact TOML is a compact Boolean algebra.

In fact, we get a good deal more than this.  Recall that an orthoalgebra regular iff every pairwise compatible subset is contained in a Boolean sub-orthoalgebra. Most orthoalgebras that arise in practice, including all lattice-ordered orthoalgebras, are regular. A {\em block} in an orthoalgebra is a maximal Boolean sub-orthoalgebra.

\begin{theorem} Let $L$ be a compact, regular TOA. Then 
\begin{enumerate}
\item[(a)] Every block of $L$ is a compact Boolean algebra, as is the center of $L$;
\item[(b)]$L$ is atomic.
\end{enumerate} \end{theorem}

\begin{proof} (a) If $L$ is regular, then a block of $L$ is the same thing as a maximal pairwise compatible set. It follows from part (d) of Proposition 3.11 that every block is closed in $L$, and hence compact. It is not hard to show that in a regular TOA the center is the intersection of the blocks. Thus we also have that $\C(L)$ is also closed, hence compact. Proposition 3.9 now supplies the result.  

To prove (b), suppose $a \in L$.  By Zorn's Lemma, there is some block $B \subseteq L$ with $a \in B$. Since $B$ is a compact Boolean algebra, it is complete and atomic; hence, $a$ can be written as the join, $\bigvee_{B} A$, of a set $A$ of atoms in $B$. Equivalently, $a = \bigvee_{B}\{\bigoplus F | F \subseteq A, F \ \text{finite}\}$. By lemma 3.8, $L$ is orthocomplete, hence, $\bigoplus A = \bigvee_{L} \{ \bigoplus F | F \subseteq A, \ F \ \text{finite}\}$ also exists, and 
is the limit of the partial sums $\bigoplus F$, $F \subseteq A$ finite. Since each partial sum lies in $B$, and $B$ is closed, $\bigoplus A \in B$. It follows that $\bigoplus A = a$. \end{proof} 

\subsection{TOAs with Isolated Zero} In \cite{ChoGre}, it is established that any TOML with an isolated point is discrete. In particular, a compact TOML with an isolated point is finite. As the example of $L(\H)$ illustrates, this is not generally true for compact lattice-ordered TOAs. This does not hold for lattice-ordered TOAs generally. Indeed, 
if $\H$ is a finite-dimensional Hilbert space, then $L(\H)$ is a compact lattice-ordered TOA in which $0$ is isolated. On the other hand, 

This is generally not true even for compact lattice-ordered TOAs. This is illustrated by the example of $L(\H)$ where $\H$ is finite-dimensional. Here we have a compact, but certainly not discrete, lattice-ordered TOA in which $0$ is isolated.  On the other hand, as we now show, compact TOAs such in which $0$ is isolated do have quite special properties. We begin with an elementary but important observation. Call an open set in a TOA space {\em totally non-orthogonal} if it contains no two orthogonal elements.

\begin{proposition} Every non-zero element of a TOA has has a totally non-orthogonal open neighborhood.\end{proposition}

\begin{proof} Let $L$ be a TOA. If $a \not = 0$, then $(a,a) \not \in \perp$. Since the latter is closed in $L^{2}$, we can find open sets $U$ and $V$ with $(a,a) \in U \times V$ and $(U \times V) \cap \perp = \emptyset$. The set $U \cap V$ is a totally non-orthogonal open neighborhood of $a$. \end{proof}

\begin{proposition} Let $L$ be a compact TOA with $0$ isolated. Then
\begin{enumerate}
\item[(a)] $L$ is atomic and of finite height;
\item[(d)] The set of atoms of $L$ is open.
\end{enumerate} \end{proposition}

\begin{proof}
(a) We first show that there is a finite upper bound on the size of a pairwise orthogonal set. Since $0$ is isolated in $L$, $L \setminus \{0\}$ is compact.
By Lemma 3.13, we can cover $L \setminus \{0\}$ by finitely many
totally non-orthogonal open sets $U_{1},...,U_{n}$. A pairwise-orthogonal
subset of $L \setminus \{0\}$ can meet each $U_{i}$ at most once, and so, can have at most $n$ elements. Now given a finite chain $x_1 < x_2 < ... < x_m$ in $L$, construct 
a pairwise orthogonal set $y_1,...,y_{m-1}$ defined by   
$y_{1} = x_{1}$ and $y_{k} = x_{k+1} \ominus y_{k}$ for $k = 2,...,m-1$. Hence, 
$m-1 \leq n$, so $m \leq n+1$. This shows that $L$ has finite height, from which it follows that $L$ is atomic. 

(b) Note that if $A$ and $B$ are any closed subsets
of $L$, then $(A \times B) \cap \perp$ is a closed, hence compact, subset of
$\perp$. Hence, the set 
\[A \oplus B := \{ a \oplus b | a \in A, b \in B \ \text{and} \ a \perp b\}
= \oplus ((A \times B) \cap \perp)\] is closed. Now note that the set of non-atoms is
precisely $(L \setminus \{0\}) \oplus (L \setminus \{0\})$. Since $0$ is isolated,
$(L \setminus \{0\})$ is closed. Thus, the set of non-atoms is closed. 
\end{proof}

\begin{remark} Notice that both the statements and the proofs of Lemma 3.13 and part (a) of Proposition 3.14 apply verbatim to any topological orthoposet, i.e., any ordered space having a closed order and equipped with a continuous orthocomplementation.
\end{remark}

If $a$ belongs to the center of a TOA $L$, then $[0,a] \times [0,a'] \subseteq \perp$. Hence, the natural isomorphism $\phi : [0,a] \times [0,a'] \rightarrow L$ given by $(x,y) \mapsto x \otimes y$ is continouous. If $L$ is compact, then so are $[0,a]$ and $[0,a']$; hence, $\phi$ is also an homeomorphism.  Since the center of an orthoalgebra is a Boolean sub-orthoalgebra of $L$, and since a Boolean algebra of finite height is finite, Proposition 3.12 has the following 

\begin{corollary} Let $L$ be a compact TOA with $0$ isolated. Then the center of $L$ is finite. In particular, $L$ decomposes, both algebraically and topologically, as the     product of finitely many compact simple TOAs.\end{corollary}

\section{Stably Ordered Topological Orthoalgebras}

In this section we consider a particularly tractable, but still quite broad, class of TOAs.

\begin{definition} We shall call an ordered topological space $L$ {\em stably ordered} iff, for every open set $U \subseteq L$, the upper-set $U\uparrow = \{ b \in L | \exists a \in U a \leq b\}$ is again open.\footnote{The term used by Priestley [13] is ``space of type $I_{i}$."} \end{definition}

\begin{remark} Note that this is equivalent to saying that the second projection mapping $\pi_2 : \leq \rightarrow L$ is an open mapping, since for open sets $U, V \subseteq L$,
\[\pi_2(( U \times V) \cap \leq) = U\uparrow \cap V.\] 
Note, too, that if $L$ carries a continuous orthocomplementation $'$, then $L$ is stably ordered iff $U\downarrow = \{ x | \exists y \in U, x \leq y\}$ be open for all open sets $U \subseteq L$. \end{remark}

\begin{example}  The following example (a variant of Example 3.3) shows that a TOA need not be stably ordered. Let $L = [0,1/4] \cup [3/4,1]$ with its usual topology, but without its usual order.
For $x, y \in L$, set $x \perp y$ iff $x + y = 1$ or $x = 0$ or $y = 0$. In any of these cases, define
$x \oplus y = x + y$. As is easily checked, this is a compact lattice-ordered TOA. However, for the clopen set $[0,1/4]$ we have $[0,1/4]\uparrow = [0,1/4] \cup \{1\}$, which is certainly not open. \end{example}

Such examples notwithstanding, most of the orthoalgebras that arise ``in nature" do seem to be stably ordered. The following is mentioned (without proof) in [13]:

\begin{lemma} Any topological $\wedge$-semilattice -- in particular, any topological lattice --  is stably ordered.\end{lemma}

\begin{proof} If $L$ is a topological meet-semilattice and $U \subseteq L$ is open, then
\[U\uparrow = \{ \ x \in L \ | \ \exists y \in U \ x \wedge y \in U\} = \pi_{1} (\wedge^{-1}(U))\]
where $\pi_{1} : L \times L \rightarrow L$ is the (open) projection map on the first factor and $\wedge : L \times L \rightarrow L$ is the (continuous) meet operation. \end{proof}

Many orthoalgebras, including projection lattices, can be embedded in ordered abelian groups. Indeed, suppose $G$ is an ordered abelian group. If $e > 0$ in $G$, let $[0,e]$ denote the set of all elements $x \in G$ with $0 \leq x \leq e$. We can endow $[0,e]$ with the following partial-algebraic structure: for $x, y \in [0,e]$, set $x \perp y$ iff $x + y \leq e$, in which case let $x \oplus y = x + y$. Define $x' = e - x$. Then $([0,e],\oplus,',0,e)$ is an effect algebra -- that is, it satisfies all of the axioms for an orthoalgebra save possibly the condition that $x \perp x$ only for $x = 0$. By a {\em faithful sub-effect algebra} of $[0,e]$, we mean a subset $L$ of $[0,e]$, containing $0$ and $e$, that is closed under $\oplus$ (where this is defined) and under $'$, {\em and} such that, for all $x, y \in L$, $x \leq y$ iff $\exists z \in L$ with $y = x + z$. 

By way of example, let $L = L(\H)$, the projection lattice of a Hilbert space $\H$, regarded as an orthoalgebra, and let $G = {\cal B}_{sa}(\H)$, the ring of bounded self-adjoint operators on $\H$, ordered in the usual way. Then $L$ is a faithful sub-effect algebra of $[0,\1]$, where $\1$ is the identity operator on $\H$. This follows from the fact that, for projections $P, Q \in L(\H)$, $P + Q \leq \1$ iff $P \perp Q$, and the fact that if $P \leq Q$ as positive operators, then $Q - P$ is a projection. 
 
\begin{lemma} Let $L$ be an orthoalgebra, let $G$ be any ordered topological abelian group with a closed cone (equivalently, a closed order), and suppose that $L$ can be embedded as a sub-effect algebra of $[0,e]$, where $e > 0$ in $G$. Then $L$, in the 
topology inherited from $G$, is a stably ordered TOA. \end{lemma}

\begin{proof} We may assume that $L$ is a subspace of $[0,e]$. Since $x \perp y$ in $L$ iff $x + y \leq e$, we have $\perp = +^{-1}([0,e]) \cap L$, which is relatively closed in $L$. The continuity of $\oplus$ and $'$ are automatic.
Suppose now that $U \cap L$ is a relatively open subset of $L$. Then, since $L$ is a faithful sub-effect algebra of $[0,e]$, the upper set generated by $U \cap L$ in $L$ is $U \uparrow \cap L$, where $U\uparrow$ is the upper set of $U$ in $[0,e]$. It suffices to show that this last is open. But $U \uparrow = \bigcup_{y \in G_{+}} U + y$, which is certainly open. \end{proof}

In particular, it follows that the projection lattice $L(\H)$ of a Hilbert space $\H$ is stably ordered in its norm topology. 

\begin{example} A {\em state} on an orthoalgebra $(L,\oplus)$ is a mapping $f : L \rightarrow [0,1]$ such that $f(1) = 1$ and, for all $a, b \in L$,  $f(a \oplus b) = f(a) \oplus f(b)$ whenever $a \oplus b$ exists. A set $\Delta$ of states on $L$ is said to be {\em order-determining} iff $f(p) \leq f(q)$ for all $f \in \Delta$ implies $p \leq q$ in $L$. In this case the mapping $L \rightarrow {\Bbb R}^{\Delta}$ given by $p \mapsto \hat{p}$, $\hat{p}(f) = f(p)$, is an order-preserving injection. Taking $G = {\Bbb R}^{\Delta}$ in Lemma 4.5, we see that $L$ is a stably-ordered TOA in the topology inherited from pointwise convergence in $G$. As a special case, note that 
the projection lattice $L = L(\H)$ has an order-determining set of states of the 
form  form $f(p) = \langle p x, x \rangle$, where $x$ is a unit vector in $\H$.  Thus, $L(\H)$ is stably-ordered also in its weak topology.\end{example}

If $U, V \subseteq L$, let us write $U \oplus V$ for $\oplus ((U \times V) \cap \perp)$, i.e., for the set of all (existing) orthogonal sums $a \oplus b$ with $a \in U$ and $b \in V$.

\begin{lemma} A TOA is stably ordered if, and only if, for every pair of open
sets $U, V \subseteq L$, the set $U \oplus V$ is also open.\end{lemma}

\begin{proof} Let $U$ and $V$ be any two open sets in $L$. Then
\begin{eqnarray*}
U \oplus V & = & \{ c \in L | c = a \oplus b, a \in U, b \in V\}\\
& = & \{ c \in L | \exists a \in U \ a \leq c \ \text{and} \ c \ominus a \in V\} \\
& = & \pi_{2}(\ominus^{-1}(V) \cap (L \times U \uparrow)) .
\end{eqnarray*}
Now, since $L$ is stably ordered, $U\uparrow$ is open, and hence, $\ominus^{-1}(V) \cap (L \times U \uparrow)$ is
relatively open in $\leq$. But as observed above, for $L$ stably ordered, $\pi_2 : \leq \rightarrow L$ is an open mapping,
so $U \oplus V$ is open. For the converse, just note that $U \uparrow = U \oplus L$. \end{proof}

Proposition 3.12 tells us that a compact TOA $L$ with $0$ isolated is atomic and of finite height. It follows easily that every element of $L$ can be expressed as a finite orthogonal sum of atoms. Let the {\em dimension, $\dim(a)$, of an element $a \in L$ be the minimum number $n$ of atoms $x_1,...,x_n$ such that $a = x_1 \oplus \cdots \oplus x_n$. Note that $a \in L$ is an atom iff $\dim(a) = 1$. 

\begin{theorem} Let $L$ be a compact, stably-ordered TOA in which $0$ is an isolated point.
Then
\begin{enumerate}
\item[(a)] The set of elements of $L$ of a given dimension is clopen.
\item[(b)] The topology on $L$ is completely determined by that on the set of atoms. \end{enumerate} \end{theorem}

\begin{proof} We begin by noting that if $A$ and $B$ are clopen subsets of $L$, then 
$A \oplus B$ is again clopen (open, by Lemma 4.7, and closed, because the image of 
the compact set $(A \times B) \cap \perp$ under the continuous map $\oplus$). Now, since $0$ is isolated, $L \setminus \{0\}$ is clopen. Since the set of non-atoms in $L$ is exactly $(L \setminus \{0\}) \oplus (L \setminus \{0\})$, it follows that the set of atoms is clopen. Define a sequence of sets $L_{k}$, $k \in {\Bbb N}$, by setting by $L_{0} = \{0\}$, $L_{1}$ = the set of atoms of $L$,
and $L_{k+1}:= L_{k} \oplus L_{1}$. These sets are clopen, as are all Boolean combinations of them. Thus, $\{ a \in L | \dim(a) = k\}  = L_{k} \setminus ( \bigcup_{i=0}^{k-1} L_{k})$ is clopen for every $k = 0,...,\dim(L)$. This proves (a). For (b), it now suffices to show that the topology on each $L_k$ is determined by that on $L_1$. Since $L_1$ and $L_2$ are clopen, Lemma 4.7 tells us that the mapping $\oplus : ((L_k \times L_1) \cap \perp) \rightarrow L_{k_+1}$ is an open surjection, and hence, a quotient mapping. Thus, the topology on $L_{k+1}$ is entirely determined by that on $L_{k}$ and that on $L_{1}$. An easy induction completes the proof. \end{proof}

\end{document}